\documentclass[showpacs,10pt,twocolumn,prb]{revtex4-1}
\usepackage{amsmath}
\usepackage{amssymb}
\usepackage{graphics}
\usepackage{epsfig}

\begin{document}

\title{Anisotropy in transport and magnetic properties of K$_{0.64}$Fe$%
_{1.44}$Se$_{2.00}$}
\author{Hechang Lei and C. Petrovic}
\affiliation{Condensed Matter Physics and Materials Science Department, Brookhaven
National Laboratory, Upton, NY 11973, USA}
\date{\today}

\begin{abstract}
We report a study of anisotropy in transport and magnetic properties of K$%
_{0.64}$Fe$_{1.44}$Se$_{2.00}$ single crystals. The anisotropy in
resistivity is up to one order of magnitude between 1.8 K and 300 K.
Magnetic susceptibility exhibits weak temperature dependence in the normal
state with decrease in temperature with no significant anomalies. The lower
critical fields $H_{c1}$\ of K$_{0.64}$Fe$_{1.44}$Se$_{2.00}$ are only about
3 Oe and the anisotropy of $H_{c1,c}$/$H_{c1,ab}$ is about 1. The critical
currents for H$\Vert $ab and H$\Vert $c are about 10-10$^{3}$ A/cm$^{2}$,
smaller than in iron pnictides and in FeTe$_{1-x}$Se$_{x}$ and nearly
isotropic.
\end{abstract}

\pacs{74.70.Xa, 74.25.Sv, 74.25.Op, 74.25.-q}
\maketitle

\section{Introduction}

Superconductivity discovery in LaFeAsO$_{1-x}$F$_{x}$ has triggered intense
research activity that resulted in critical temperatures up to 56 K in
pnictide materials.\cite{Kamihara}$^{-}$\cite{Wen HH} Soon after, several
types of iron-based superconductors have been discovered, such as AFe$_{2}$As%
$_{2}$ (A = alkaline or alkaline-earth metals, 122-type),\cite{Rotter}$^{,}$%
\cite{Chen GF2} LiFeAs (111-type),\cite{Wang XC} (Sr$_{4}$M$_{2}$O$_{6}$)(Fe$%
_{2}$Pn$_{2}$) (M = Sc, Ti or V, 42622-type),\cite{Ogino}$^{,}$\cite{Zhu XY}
and $\alpha $-PbO type FeSe (11-type)\cite{Hsu FC} etc. The 11-type
materials FeSe, FeTe$_{1-x}$Se$_{x}$,\cite{Yeh KW} and FeTe$_{1-x}$S$_{x}$%
\cite{Mizuguchi} provided an example \ of iron based superconductivity in a
rather simple crystal structure without the charge reservoir layer. Yet,
these simple binary structures share a square-planar lattice of Fe with
tetrahedral coordination and similar Fermi surface topology with other
iron-based superconductors.\cite{Subedi} Furthermore, 11-type
superconductors contain some distinctive structural and physical features,
such as interstitial iron Fe$_{1+y}$Te and the significant pressure effect.%
\cite{Zhang LJ}$^{-}$\cite{Medvedev} Under external pressure, the $T_{c}$
can be increased from 8 K to 37 K and the $dT_{c}/dP$ can reach 9.1 K/GPa,
the highest increase in all iron-base superconductors.\cite{Medvedev} This
behavior may be understood from the observation related to the anion height
between Fe and As (or Se, Te) layers. There is an optimal distance around
1.38 \AA\ with a maximum transition temperature $T_{c}\simeq $ 55 K.\cite%
{Mizuguchi3} The anion height in FeSe decreases gradually with the pressure
increase towards the optimal value thereby increasing $T_{c}$.\cite%
{Mizuguchi3} Quite importantly, high upper critical fields and currents were
demonstrated in iron based superconductors.\cite{Yuan}$^{-}$\cite{Kida}

Another method for tuning of the anion height is the intercalation between
FeSe layers that can change both the local environment of Fe-Se tetrahedron
and the average crystal structure. The intercalation could also decrease
dimensionality of conducting bands. The presence of low energy electronic
collective modes in layered conductors helps screening Coulomb interaction,
which may contribute constructively to superconductivity.\cite{Bill} This is
seen in iron based superconductors: the $T_{c}$ increases from 11-type to
1111-type. Very recently the $T_{c}$\ was raised in iron selenide material
to about 30 K by intercalating K, Rb, Cs, and Tl between the FeSe layers
(AFeSe-122 type),\cite{Guo}$^{-}$\cite{Fang MH} as opposed to pressure. The
intercalation of alkaline metals decreases Se height,\cite{Krzton-Maziopa}
and changes the average space group from P4/nmm of FeSe to I4/mmm of
AFeSe-122 type. The Fe-Se interlayer distances are expanded and may
contribute to electronic and magnetic dimensionality reduction. Furthermore,
insulator-superconductor transition (IST) can be induced by tuning the Fe
stoichiometry in (Tl$_{1-x}$K$_{x}$)Fe$_{2-y}$Se$_{2}$ (0 $\leq $ x $\leq $
1, 0 $\leq $ y $\leq $ 1).\cite{Fang MH} This suggests that the
superconductivity of AFeSe-122 type is in proximity of a Mott insulating
state.\cite{Fang MH} Thus it is of interest to study electronic and magnetic
anisotropy in normal and superconducting states in AFeSe-122 in order to
shed more light on the superconducting mechanism and possible symmetry of
the order parameter.

In this work, we report the anisotropy in electronic transport and
magnetization in the normal state of K$_{0.64(4)}$Fe$_{1.44(4)}$Se$%
_{2.00(0)} $ single crystals. We also present anisotropic parameters of the
superconducting state.

\section{Experiment}

Single crystals of K$_{x}$Fe$_{2}$Se$_{2}$ were grown by self-flux method
with nominal composition K$_{0.8}$Fe$_{2}$Se$_{2}$. Prereacted FeSe and K
pieces (purity 99.999\%, Alfa Aesar) were put into the alumina crucible, and
sealed into the quartz tube with partial pressure of argon. The quartz tube
was heated to 1030 $%
{{}^\circ}%
C$, kept at this temperature for 3 hours, and then slowly cooled to 730 $%
{{}^\circ}%
C$ with 6 $%
{{}^\circ}%
C$/hour. Plate-like crystals up to 5$\times $5$\times $1 mm$^{3}$ can be
grown. X-ray diffraction (XRD) spectra were taken with Cu K$_{\alpha }$
radiation ($\lambda $ = $1.5418$ \AA ) using a Rigaku Miniflex X-ray
machine. The lattice parameters were obtained by fitting the XRD spectra
using the Rietica software.\cite{Hunter} The elemental analysis was
performed using an energy-dispersive x-ray spectroscopy (EDX) in an JEOL
JSM-6500 scanning electron microscope. Electrical resistivity $\rho (T)$
measurements were performed in Quantum Design \ PPMS-9. The in-plane
resistivity $\rho _{ab}(T)$ was measured using a four-probe configuration on
rectangularly shaped and polished single crystals with current flowing in
the ab-plane of tetragonal structure. The c-axis resistivity $\rho _{c}(T)$
was measured by attaching current and voltage wires on the opposite sides of
the plate-like sample.\cite{Wang XF}$^{,}$\cite{Edwards}\textbf{\ }Since the
sample surface is easily oxidized, sample manipulation in air was limited to
10 minutes. Sample dimensions were measured with an optical microscope Nikon
SMZ-800 with 10 um resolution. Electrical transport and heat capacity
measurements were carried out in PPMS-9 from 1.8 to 300 K. Magnetization
measurements were performed in a Quantum Design Magnetic Property
Measurement System (MPMS) up to 5 T.

\section{Results and Discussion}

\begin{figure}[tbp]
\centerline{\includegraphics[scale=0.8]{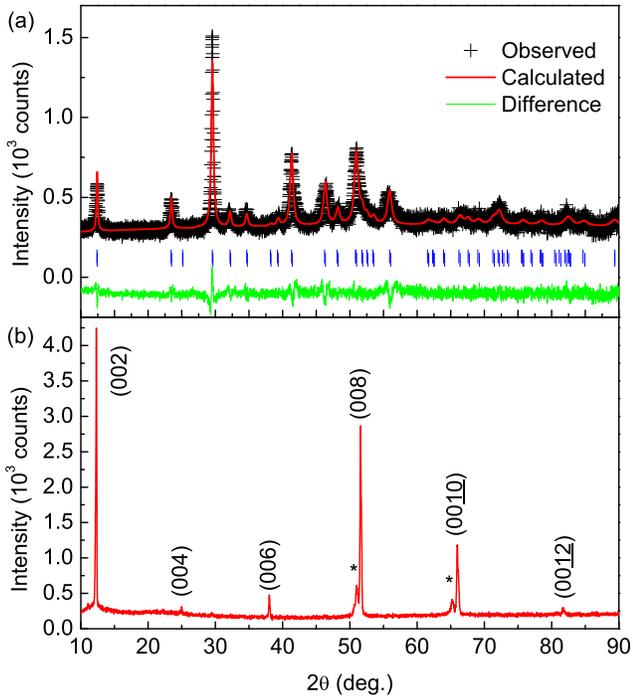}} \vspace*{-0.3cm}
\caption{(a) Powder and (b) single crystal XRD patterns of K$_{x}$Fe$_{2-y}$%
Se$_{2}$, respectively.}
\end{figure}

Fig. 1(a) shows the X-ray diffraction (XRD) results of the ground crystal.
It confirms phase purity with no extrinsic peaks. The powder pattern can be
indexed in the I4/mmm space group with fitted lattice parameters a =
0.39109(2) nm, c = 1.4075(3) nm ($R_{p}$ = 2.272 and $R_{wp}$ = 3.101),
consistent with reported results.\cite{Guo}$^{-}$\cite{Ying JJ} On the other
hand, XRD spectra of a single crystal reveal that the crystal surface is
normal to the c axis with the plate-shaped surface parallel to the ab-plane
(Fig. 1(b)).\ Furthermore, there is another weak series of (00l) diffraction
peaks (labeled by the asterisks) associated with the main peaks. These two
distinct sets of reflections could arise from the inhomogeneous distribution
of K atoms, which also observed in other AFeSe-122 single crystals.\cite%
{Wang HD}$^{,}$\cite{Luo XG} The average stoichiometry was determined from
EDX by examination of multiple points on the crystals. The measured
compositions are K$_{0.64(4)}$Fe$_{1.44(4)}$Se$_{2.00(0)}$ (noted as K$_{x}$%
Fe$_{2-y}$Se$_{2}$), indicating substantial with of formation and the
existence of K and Fe vacancies. We also measured the composition mapping
using EDX. The results exhibit that the spatial distribution of K, Fe, and
Se are homogenous.

\begin{figure}[tbp]
\centerline{\includegraphics[scale=0.8]{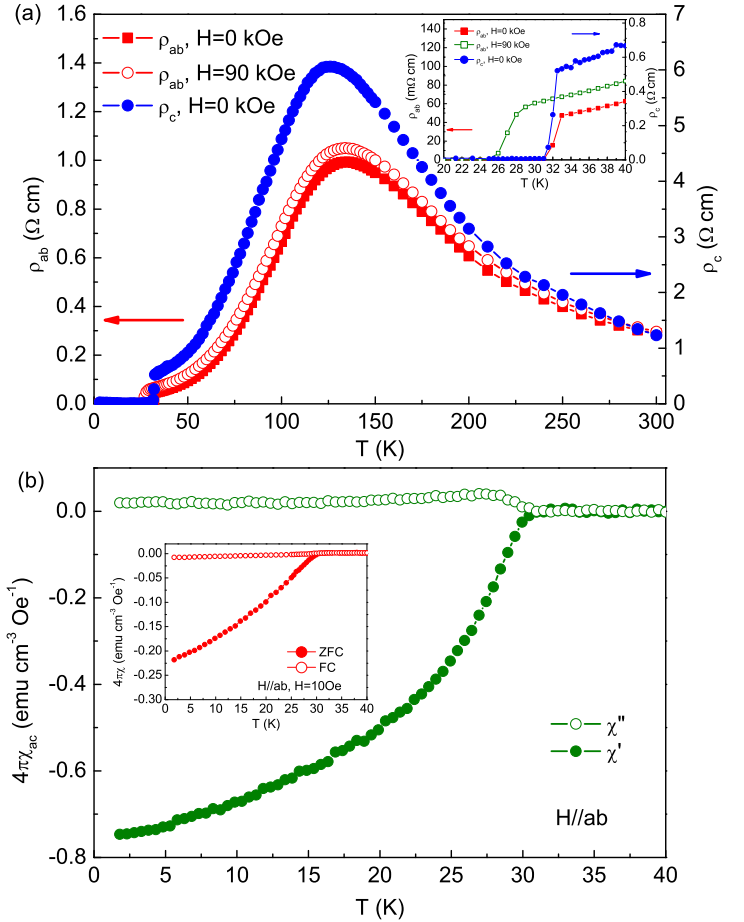}} \vspace*{-0.3cm}
\caption{(a) Temperature dependence of the resistivity $\protect\rho _{ab}$%
(T) and $\protect\rho _{c}$(T) of K$_{x}$Fe$_{2-y}$Se$_{2}$ with and without
H=90 kOe along c axis. Inset: enlarged resistivity curve near T$_{c}$. (b)
Temperature dependence of ac magnetic susceptibility of K$_{x}$Fe$_{2-y}$Se$%
_{2}$ in H$_{ac}$=1 Oe. Inset: temperature dependence of dc magnetic
susceptibility with ZFC and FC.}
\end{figure}

The main panel of Fig. 2(a) shows the temperature dependence of resistivity
in zero field from 1.9 K to 300 K for current along ab plane and c axis. At
higher temperatures, both $\rho _{ab}$(T) and $\rho _{c}$(T) show a
metal-insulator transition with a maximum resistivity at about 135 K and 125
K, respectively. The resistivity maximum ($\rho _{\max }$) might be related
to a scattering crossover arising from a structure or magnetic phase
transition and is a typical behavior in AFeSe-122 system.\cite{Guo}$^{-}$%
\cite{Fang MH} Furthermore, the temperature dependence of the $\rho _{\max }$
depends on the extent of iron deficiency in the crystals.\cite{Wang DM} From
our resistivity data in the normal state below 300 K, the ratio $\rho _{c}$/$%
\rho _{ab}$ is about 4-12. The anisotropy is much smaller than in (Tl,K)$%
_{x} $Fe$_{2-y}$Se$_{2}$ and (Tl,Rb)$_{x}$Fe$_{2-y}$Se$_{2}$ system, where $%
\rho _{c}$/$\rho _{ab}$ = 70-80 and 30-45.\cite{Fang MH}$^{,}$\cite{Wang HD}
On the other hand, both of $\rho _{ab}$(T) and $\rho _{c}$(T) undergo a very
sharp superconducting transition at $T_{c,onset}$ = 33 K, shown in the inset
of Fig. 2(a). At 90 kOe, the resistivity transition width is broader and the
onset of superconductivity shifts to 28 K. However, the $\rho _{\max }$
curve has no obvious shift in magnetic field up to 90 kOe for current
transport along both crystallographic axes.

Fig. 2(b) shows the temperature dependence of the ac susceptibility of K$%
_{x} $Fe$_{2-y}$Se$_{2}$ single crystal with H$\parallel $ab. A clear
superconducting transition appearers at T = 31 K. This is consistent with
the resistivity results. The superconducting volume fraction is about 75\%
at 1.8 K, indicating the bulk superconductivity in the sample. The broad
transitions in $\chi ^{\prime }$ and $\chi ^{"}$ point to microscopic
inhomogeneity. Inset in Fig. 2(b) shows the dc magnetic susceptibility for H$%
\parallel $ab with zero-field cooling (ZFC) and field cooling (FC).
Diamagnetism can be clearly observed in both measurement and the $%
T_{c,onset} $ is almost the same as that determined from the ac
susceptibility. On the other hand, the magnetization measured with FC is
very small, which is a common behavior in two-dimensional superconductors,
such as (Pyridine)$_{1/2}$TaS$_{2}$,\cite{Prober} and Ni$_{x}$TaS$_{2}$.\cite%
{Li LJ} The small magnetization values for FC is likely due to the
complicated magnetic flux pinning effects in the layered compounds.\cite%
{Prober}

Temperature dependence of magnetic susceptibility in the normal state is
shown in Fig. 3(a) for H$\Vert $ab and H$\Vert $c with H = 1 kOe. A sudden
drop at about 30 K corresponds to the superconducting transition. For H$%
\Vert $c, $\chi _{c}$ weakly decreases with temperature below 300 K and
exhibits a weak upturn below 120 K. When the magnetic field is in the ab
plane, $\chi _{ab}$ exhibits similar behavior but the minimum of
susceptibility is located at about 175 K. The magnetic susceptibility
enhancement with increase in temperature above 200 K is neither Pauli nor
Curie-Weiss. It suggests the presence of magnetic interactions. This has not
only been observed in other AFeSe-122 compounds,\cite{Ying JJ}$^{,}$\cite%
{Wang AF}$^{,}$\cite{Fang MH}$^{,}$\cite{Lei HC} but also in BaFe$_{2}$As$%
_{2}$ due to two dimensional short range AFM spin fluctuations.\cite{Zhang
GM}$^{,}$\cite{Matan} The AFM interaction is possibly related to the Fe
deficiency and is an intrinsic properties of AFe$_{2}$Ch$_{2}$ (Ch = S, Se).

\begin{figure}[tbp]
\centerline{\includegraphics[scale=0.45]{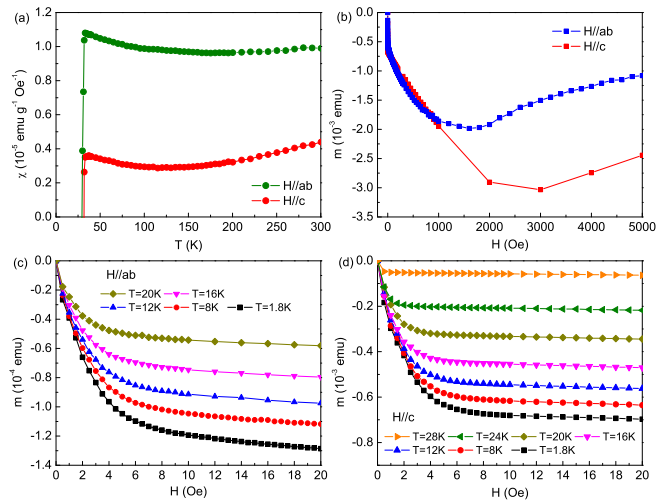}} \vspace*{-0.3cm}
\caption{(a) Temperature dependence of magnetic susceptibility measured at 1
kOe for K$_{x}$Fe$_{2-y}$Se$_{2}$ crystal with H$\Vert $ab and H$\Vert $c.
(b) Magnetization curve of K$_{x}$Fe$_{2-y}$Se$_{2}$ at T = 1.8 K for H$%
\Vert $ab and H$\Vert $c. (c) and (d) Low field parts of m(H) at various
temperature for H$\Vert $ab and H$\Vert $c, respectively.}
\end{figure}

The initial dc magnetization versus field m(H) at T = 1.8 K for both
directions is shown in Fig. 3(b). The shape of the m(H) curves points that K$%
_{x}$Fe$_{2-y}$Se$_{2}$ is a typical type-II superconductor. The peak in
m(H) is about 2 kOe for H$\parallel $c, consistent with the previous report.%
\cite{Guo} However, it should be noted that $H_{c1}$ is often much smaller
than the peak value in m(H) curve. In K$_{x}$Fe$_{2-y}$Se$_{2}$, the m(H)
curve deviates from linearity at much lower field. The enlarged parts are
shown in Fig. 3(c) and (d). The H$_{c1}$ is usually determined by the field
where the m(H) deviates from linear relation.\cite{Ren C} However, small $%
H_{c1}$ introduces the significant error, so it is hard to evaluate the $%
H_{c1}$(0) using $H_{c1}$(T) = $H_{c1}$(0)[1 - ($T/T_{c}$)$^{2}$]. The
approximate $H_{c1,ab}$(T = 1.8 K) and $H_{c1,c}$(T = 1.8 K) are 3.0(5) Oe.

\begin{figure}[tbp]
\centerline{\includegraphics[scale=0.45]{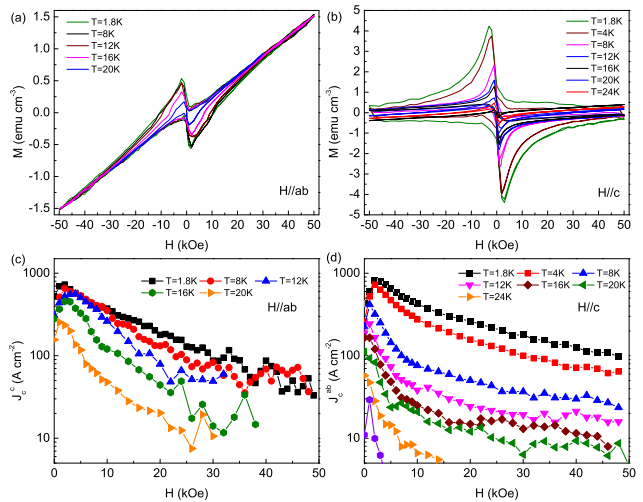}} \vspace*{-0.3cm}
\caption{Magnetization hysteresis loops of K$_{x}$Fe$_{2-y}$Se$_{2}$ for (a)
H$\Vert $ab and (b) H$\Vert $c. (c) In-plane and (d) interplane
superconducting critical currents as determined from magnetization
measurements using the Bean model.}
\end{figure}

Fig. 4(a) and (b) show the magnetization loops for H$\Vert $c and H$\Vert $%
ab with field up to 50 kOe. The paramagnetic background exists for both
directions, and is more obvious for H$\Vert $ab. This paramagnetic
background originates from the non-superconducting fraction. The shapes of
M(H) and M(T) (Fig. 3) are typical of type-II superconductors with some
electromagnetic granularity.\cite{Kupfer}$^{,}$\cite{Senatore} The critical
current is determined from the Bean model.\cite{Bean}$^{,}$\cite{Gyorgy} For
a rectangular-shaped crystal with dimension c $<$ a $<$ b, when H$\Vert $c,
the in-plane critical current density $J_{c}^{ab}(H)$ is given by

\begin{equation*}
J_{c}^{ab}(H)=\frac{20\Delta M(H)}{a(1-a/3b)}
\end{equation*}

where a and b (a$<$b) are the in-plane sample size in cm, $\Delta M(H)$ is
the difference between the magnetization values for increasing and
decreasing field at a particular applied field value, measured in emu/cm$%
^{3} $, and $J_{c}^{ab}(H)$ is the critical current in A/cm$^{2}$. It should
be noted that the paramagnetic background has no effect on the calculation
of $\Delta M(H)$. The situation is more complex when H$\Vert $ab. There are
two different current densities: one is the vortex motion across the planes,
$J_{c}^{c}(H)$, and another is parallel to the planes, $J_{c}^{\Vert }(H)$.
Usually $J_{c}^{\Vert }(H)\neq J_{c}^{ab}(H)$ If assuming $a,b\gg c/3\cdot
J_{c}^{\Vert }(H)/J_{c}^{c}(H)$,\cite{Gyorgy} we can obtain $%
J_{c}^{c}(H)\approx 20\Delta M(H)/c$. Magnetic field dependence of $%
J_{c}^{c}(H)$ and $J_{c}^{ab}(H)$ is shown in Fig. 4(c) and (d). It can be
seen that the critical current decreases with applied field and the ratio of
$J_{c}^{c}(H)$/$J_{c}^{ab}(H)$ is approximately 1. The critical current
densities for both directions are 10-10$^{3}$ A/cm$^{2}$, which is much
smaller than those of BaFe$_{2-x}$Co$_{x}$As$_{2}$ in the same temperature
range.\cite{Tanatar}

\section{Conclusion}

In summary, we have presented anisotropic transport and magnetic properties
of K$_{0.64(4)}$Fe$_{1.44(4)}$Se$_{2.00(0)}$ single crystals with $%
T_{c,onset}$ = 33 K and free of iron impurities. The resistivity anisotropy
is much smaller than in other AFeSe-122 compounds. Magnetization decreases
in the normal state with decreasing temperature from 300 K, which suggest
that the presence of AFM interactions. The lower critical fields H$_{c1}$\
are only about 3 Oe at 1.8 K and the anisotropy of $H_{c1,c}$/$H_{c1,ab}$ is
about 1. The critical current values are isotropic and is about 10-10$^{3}$
A/cm$^{2}$ for both directions below 50 kOe.

\section{Acknowledgements}

We thank John Warren for help with SEM measurements. Work at Brookhaven is
supported by the U.S. DOE under Contract No. DE-AC02-98CH10886 and in part
by the Center for Emergent Superconductivity, an Energy Frontier Research
Center funded by the U.S. DOE, Office for Basic Energy Science.

\end{document}